\documentclass[12pt]{article}

\usepackage{graphicx}
\usepackage{amsmath,amssymb}
\usepackage{bm}
\usepackage{graphicx, color}
\usepackage{wrapfig}
\begin{document}
\title{
\begin{flushright}
\ \\*[-80pt] 
\begin{minipage}{0.2\linewidth}
\normalsize
\end{minipage}
\end{flushright}
{\Large \bf 
A Model connecting 
Quantum, Diffusion, Soliton, and Periodic Localized States 
under Brownian motion
\\*[20pt]}}

\author{
\centerline{
Hajime~Isimori$^{1,}$\footnote{E-mail address: ishimori@gauge.scphys.kyoto-u.ac.jp}   }
\\*[20pt]
\centerline{
\begin{minipage}{\linewidth}
\begin{center}
$^1${\it \normalsize
Department of Physics, Kyoto University, 
Kyoto 606-8502, Japan}  
\end{center}
\end{minipage}}
\\*[50pt]}

\date{
\centerline{\small \bf Abstract}
\begin{minipage}{0.9\linewidth}
\medskip 
\medskip 
\small
 We  propose new equations of motion under the theory of 
the Brownian motion to connect the states of 
quantum, diffusion, soliton, and periodic localization. 
The new equations are nothing but the classical equations of motion 
with two additional terms and the one of them can be regarded as the 
the quantum potential. By choosing a parameter space, 
various important states are obtained. 
Further, the equations contain other interesting phenomena 
such as general dynamics of 
diffusion process, collapse of the soliton, the 
nonlinear extension of the Schr\"dinger equation, 
and the dynamics of phase transition. 
\end{minipage}
}

\begin{titlepage}
\maketitle
\thispagestyle{empty}
\end{titlepage}

\section{Introduction}

The description of microscopic world governed by quantum mechanics 
is quite different from the one of microscopic world with classical mechanics 
\cite{Acosta,Neufelda}.  
Whereas, it is an important issue to develop a framework that 
includes both of classical and quantum physics 
for the purpose to study mixed dynamics \cite{Cohen,Berges,Nielsen}. 
Formally, one needs to handle large number of quantum particles 
to organize macroscopic world. 
The method treating quantum many-body system 
is accurate but time-consuming to search a numerical solution 
unless one can find a good approximation for individual circumstances. 
On the other hand, extension of classical equations of motion 
by adding quantum potential which yields the Schr\"odinger equation 
is much simpler to involve classical-quantum dynamics \cite{Site,Markowich,Chou}. 
One natural motivation of the quantum potential approach is 
to explain macroscopic quantum phenomena \cite{Appleby,Matzkin}. 
For the macroscopic level, classical equations of motion 
may be applicable even for quantum phenomena. 
In this paper, we propose more general equations 
under the theory of the Brownian motion 
as the origin of quantum potential to connect 
some famous equations of classical and quantum physics. 
As a result of this connection, quantum, diffusion, soliton, and periodic localized
states can be derived by solving the general equations of motion.

Basic concept is to generalize the idea of the 
Nelson's stochastic equations \cite{Nelson}. 
These are based on the Langevin equations and the Fokker-Planck equations 
to describe non-differentiable process. They can lead the 
Schr\"odinger equation by assuming some conditions 
such as the universal relation and the specific form of the mean acceleration. 
We generalize the Langevin equations so that 
both of the Nelson equations and the Schr\"odinger equation 
are extended. Including the nonlinear Schr\"odinger equation (NLS), 
we provide several new equations that contain interesting solutions.

The extension of the Nelson equations was already attempted in the literature. 
In the paper \cite{Davidson}, Davidson generalized the equations 
to remedy the universal relation: original relation is 
$\hbar=2mD$ and modified one is 
$\hbar=\sqrt{1-2s}mD$, where $m$ is a particle mass, 
$D$ is the diffusion coefficient, and $s$ is a free parameter. 
This modification not only reconstruct a hypothesis of the relation 
but also it is possible to make physical continuation to 
the diffusion equation by assuming $1-2s<0$.  
Moreover, as is known that the limit $\hbar\rightarrow 0$ recovers 
classical physics, choices of the parameter as 
$s=1/2-\hbar^2/2m^2D^2$, $s=1/2$, and $s>1/2$  
represent quantum, classical, and diffusion equations, 
respectively. Meanwhile, there is another extension of 
the Nelson equations with a different motivation \cite{Isimori}. 
We found a way to put them together in a 
framework of generalized equations of motion.

The purpose of our generalization is to seek a proof of 
quantum potential in a similar way of the work \cite{Nottale}. 
With the modified Nelson equations, 
new static solutions are easily obtained. 
For instance, the soliton solution in our model 
is one of a simple prediction because 
it can be distinguished from other soliton solutions (see e.g. \cite{Gerdjikov})
by comparing the density function. 
Besides, we also predict new dynamics 
such as a generalized diffusion process by inventing 
new diffusion equations. In several points, 
the existence of quantum potential in the origin of 
the Brownian motion can be examined.

This paper is organized as follows. 
In section 2, we present generalized equations of motion 
with the generalization of Nelson's equations including a brief 
review. In section 3, the Schr\"odinger equation is derived from the 
equations of motion with the extension to a Hilbert space. 
Section 4 deals with the diffusion equation by choosing a 
parameter space of the equations proposed in section 2. 
Section 5 leads the soliton equation to exhibit new type of soliton solution. 
In section 6, we obtain the periodic localized solution that 
exists in other parameter space of generalized equations of motion. 
Section 7 is devoted to the summary.

\section{Modified Nelson equations}

The Langevin equation can describe  
the Brownian motion with a random movement. 
Normally, this equation computes a motion of 
differentiable process. Suppose $\bm x(t)$ 
is a stochastic process, mean forward derivative is 
given by ${\cal D}_+\bm x(t)
=\lim_{\Delta t\rightarrow +0}
\langle (\bm x(t+\Delta t)-\bm x(t))/\Delta t\rangle$ 
and mean backward derivative is 
${\cal D}_-\bm x(t)=\lim_{\Delta t\rightarrow +0}
\langle (\bm x(t)-\bm x(t-\Delta t))/\Delta t\rangle$, 
where $\langle\rangle$ represents mean value. 
For differentiable process, mean derivatives are equal 
${\cal D}_+\bm x(t)={\cal D}_-\bm x(t)$. 
However, there are many important processes which 
are not differentiable. To extend the equation to apply non-differentiable 
processes, independent equation is set for mean forward and 
backward velocities individually. 
Basic procedure is formalized by Nelson in \cite{Nelson}.

Let us consider kinematics of Markoff process. 
The position $\bm x(t)$ of the Brownian 
particle satisfies
\begin{eqnarray}
d_+\bm x=\bm b_+(t,\bm x(t)) dt+d\bm w,
\quad
d_-\bm x=\bm b_-(t,\bm x(t)) dt+d\bm w,
\end{eqnarray}
where $\bm b_+$ ($\bm b_-$) is the mean 
forward (backward) velocity and 
$\bm w$ describes the Wiener process 
which is Gaussian with mean zero. 
For the Brownian motion, it is usually set 
$\langle dw_i dw_j\rangle=2D\delta_{ij}dt$. 
The Langevin equations provide 
the space-time evolution of $\bm b_\pm$. 
Here we assume modified differential equations:
\begin{eqnarray}
\begin{split}
\label{MNEa}
d_+ \bm b_-
&=s_1(d_+-d_-)\bm b_--s_2d_+d_+\bm b_-
-\beta \bm vdt+\bm K(\bm x(t))dt
+d\bm B(t),
\\
d_- \bm b_+
&=-s_1(d_+-d_-)\bm b_++s_2d_-d_-\bm b_+
+\beta \bm vdt+\bm K(\bm x(t))dt
+d\bm B(t),
\end{split}
\end{eqnarray}
where $m\beta\bm v$ is the force of friction, 
$m\bm K$ is a body force ($m\bm K=-\nabla V$), 
$\bm B$ is a Wiener process representing the residual 
random impacts, and $d\bm B$ is Gaussian with mean zero. 
With free parameters $s_1$ and $s_2$, 
the Nelson equations are extended. 

Suppose $\bm b_\pm(t,\bm x)$ can be expanded in 
Taylor series, we get
\begin{eqnarray}
\begin{split}
d_+\bm b_-
&=dt\partial_t\bm b_-
+((\bm b_+dt+d\bm w)\cdot\nabla )\bm b_-
+\frac12(d\bm w)_i(d\bm w)_j
\partial_i\partial_j\bm b_-,
\\
d_-\bm b_+
&=dt\partial_t\bm b_+
+((\bm b_-dt+d\bm w)\cdot\nabla )\bm b_+
-\frac12(d\bm w)_i(d\bm w)_j
\partial_i\partial_j\bm b_+,
\\
d_+ d_+ \bm b_-
&=(d\bm w)_i(d\bm w)_j
\partial_i\partial_j\bm b_-,
\quad
d_- d_- \bm b_+
=(d\bm w)_i(d\bm w)_j
\partial_i\partial_j\bm b_+,
\end{split}
\end{eqnarray}
where order $(dt)^2$ is neglected. 
Taking the average of Eqs. (\ref{MNEa}), 
our generalized Langevin equations read
\begin{eqnarray}
\begin{split}
\label{NSE0}
(\partial_ t+\bm b_+\cdot\nabla
+D\nabla^2-s_1(\bm b_+-\bm b_-)\cdot\nabla
-2D(s_1-s_2)\nabla^2)\bm b_-dt
&=-\beta \bm vdt+\bm K(\bm x(t))dt,
\\
(\partial_ t+\bm b_-\cdot\nabla
-D\nabla^2+s_1(\bm b_+-\bm b_-)\cdot\nabla
+2D(s_1-s_2)\nabla^2)\bm b_+dt
&=\beta \bm vdt+\bm K(\bm x(t))dt.
\end{split}
\end{eqnarray}
Dividing by $dt$ and summing the two equations, we obtain
\begin{eqnarray}
\partial_ t\bm v+\bm v\cdot\nabla\bm v
-\bm u\cdot\nabla\bm u
-D\nabla^2\bm u
+2s_1\bm u\cdot\nabla\bm u
+2D(s_1-s_2)\nabla^2\bm u
=\bm K,
\end{eqnarray}
where $\bm v=(\bm b_++\bm b_-)/2$ is the current velocity and 
$\bm u=(\bm b_+-\bm b_-)/2$ is the osmotic velocity. 
Further, we assume the velocities satisfy the 
Fokker-Planck equations
\begin{eqnarray}
\label{FPE}
\frac{\partial\rho}{\partial t}
+\nabla\cdot(\rho\bm v)=0,
\quad
\bm u-D\nabla\ln\rho=0.
\end{eqnarray}
Now, we get equations 
that describe time dependence of $\bm u$ and $\bm v$:
\begin{eqnarray}
\begin{split}
\label{MNS}
\frac{\partial\bm u}{\partial t}
&=-D\nabla^2\bm v
-\nabla(\bm v\cdot\bm u),
\\
\frac{\partial\bm v}{\partial t}
&=-\frac12\nabla \bm v^2
+\frac{1-2s_1}{2}\nabla\bm u^2
+(1-2s_1+2s_2)D\nabla^2\bm u
-\frac1m \nabla V,
\end{split}
\end{eqnarray}
here and hereafter, we assume the current velocity 
has no rotation. 
For a case $s_1=s_2=0$, they coincide with the Nelson's equations, 
for $s_2=0$, they become the Davidson's equations \cite{Davidson}, and 
for $s_1=s_2=1/2$, they become the equations 
we proposed in previous paper \cite{Isimori}. 
The second equation is derived by the sum of two equations (\ref{NSE0}), 
then there is another independent equation. 
The subtraction of Eqs. (\ref{NSE0}) yields
\begin{eqnarray}
\label{vis}
2(s_1-1)(\bm u\cdot\nabla)\bm v
+D(2s_1-2s_2-1)\nabla^2\bm v
=\beta\bm v.
\end{eqnarray}
This equation will determine the magnitude of friction coefficient 
$m\beta$. When $\beta=0$ and $\bm v$ is constant, 
the equation becomes trivial. 
In general, right-hand side (RHS) may not be proportional to 
the current velocity, then, we shall need an extension 
$\beta v_i\rightarrow \beta_{ij}v_j$ by modifying Eqs. (\ref{MNEa}).

\section{Quantum process}
The original concept of Nelson's idea is to lead the 
Schr\"odinger equation from Eqs. (\ref{MNS}). 
For convenience, we first consider 
the Lagrangian and the Hamiltonian from the equations of motion. 
According to \cite{Cresson}, they are given by
\begin{eqnarray}
\begin{split}
L&=\frac12m\bm v^2
+\frac{mD^2}{2}(1-2s_1)(\nabla\ln\rho)^2
+mD^2(1-2s_1+2s_2)\nabla^2\ln\rho
-V,
\\
H&=\frac12m\bm v^2
-\frac{mD^2}{2}(1-2s_1)(\nabla\ln\rho)^2
-mD^2(1-2s_1+2s_2)\nabla^2\ln\rho
+V.
\end{split}
\end{eqnarray}
When the Hamiltonian is constant written by $E$, 
the current velocity is zero, $s_2=0$, 
and $f(\bm x)=\sqrt{\rho(\bm x)}$, 
we get the equation 
\begin{eqnarray}
\label{ene}
E
=-{2(1-2s_1)mD^2}
\frac{\nabla^2 f}{f}+V.
\end{eqnarray}
It looks like the Schr\"odinger equation if 
$2(1-2s_1)mD^2$ is constant and corresponds 
to $\hbar^2/2m$. 
If we use $\rho$ instead of $f$, the first term of RHS corresponds 
to the quantum potential. 
Since $\rho$ is real, 
the function $f$ is also real.  
It approaches more to the Schr\"odinger 
equation if we consider the time dependence 
of $\rho$ and $\bm v$. 
Assuming $\bm v$ is gradient of some function, 
given by $D'\nabla w$ with constant $D'$, and $s_2=0$, 
Eqs. (\ref{MNS}) can be rewritten as
\begin{eqnarray}
\begin{split}
D\frac{\partial(\ln\rho)}{\partial t}
&=-DD'\nabla^2 w
-DD'\nabla w\cdot\nabla\ln\rho,
\\
D'
\frac{\partial w}{\partial t}
&=-\frac12 \bm v^2
+\frac{1-2s_1}2\bm u^2
+(1-2s_1)D^2\nabla^2 (\ln\rho)
-\frac1m  V.
\end{split}
\end{eqnarray}
Giving a new function $f=e^{\frac12 \ln\rho+iaw}$, 
we have
\begin{eqnarray}
\nabla^2 f
=(\frac14\nabla\ln\rho\cdot\nabla\ln\rho
-a^2\nabla w\cdot\nabla w
+\frac12\nabla^2\ln\rho)f
+ia(\nabla\ln\rho\cdot\nabla w+\nabla^2w)f
\end{eqnarray}
If we take $a=D'/2D\sqrt{1-2s_1}$, 
it becomes
\begin{eqnarray}
\nabla^2 f
=\frac{1}{2(1-2s_1)D^2}
(D'\frac{\partial w}{\partial t}
+\frac{V}{m})f
-\frac{ia}{D'}\frac{\partial\ln\rho}{\partial t}f.
\end{eqnarray}
Then, this can be
\begin{eqnarray}
\label{MNE}
2i\sqrt{1-2s_1}mD\frac{\partial f}{\partial t}
=-2(1-2s_1)mD^2\nabla^2 f
+Vf,
\end{eqnarray}
Let $D=\hbar/2\sqrt{1-2s_1}m$ 
or $s_1=1/2-\hbar^2/8m^2D^2$, 
it is the same to the Schr\"odinger equation
\begin{eqnarray}
\label{SLE}
i\hbar \frac{\partial f}{\partial t}
=-\frac{\hbar^2}{2m}\nabla^2 f
+Vf.
\end{eqnarray}
Assuming $f(t,\bm x)=f(\bm x)e^{-iEt/\hbar}$ (i.e. $aw(t,\bm x)=aw(\bm x)-Et/\hbar$) 
for the case of constant energy, it is consistent 
with Eq. (\ref{ene}). 
A difference from quantum mechanics is that 
the function $f$ is scalar while the wave function 
is vector in a Hilbert space. 
The extension to a Hilbert space is necessary to 
assign quantum charges and consider multi-particle system. 
This extension is not difficult if we employ 
infinite number of probability densities and velocities. 
Taking stochastic process $\bm x^{(k)}$ for each component 
of a Hilbert space and giving mean velocities 
for each process $\bm b_+^{(k)}$ and $\bm b_-^{(k)}$,  
with $k$ runs from 1 to infinity. 
For all the velocities, we apply the modified Langevin equations 
and the Fokker-Planck equations, i.e. Eqs. (\ref{MNE}) and 
Eqs. (\ref{FPE}). Then we obtain the Schr\"odinger like equation 
for each $k$ to the function 
$f^{(k)}=e^{\frac12\ln\rho^{(k)}+ia w^{(k)}}$. 
As for the potential $V^{(k)}$, it is not necessary 
to take independent function in each component. 
That is, they can be functions of other components 
$V^{(k)}=V^{(k)}(f^{(1)},f^{(2)},\cdots)$. 
When the potential can be written in linear combinations, 
we can take $V^{(k)}f_k\rightarrow \sum V_{k\ell}f_\ell$. 
More generally, the potential will be given by 
operator system with the function of $f_k$ through other components 
so that we may write $V^{(k)}f_k\rightarrow{\hat V} f_k$. 
Rewriting $f_k=\psi$, the equations turn into
\begin{eqnarray}
i\hbar \frac{\partial }{\partial t}\psi
=-\frac{\hbar^2}{2m}\nabla^2 \psi
+\hat V\psi.
\end{eqnarray}
Now we can deal with any potential and any state of quantum mechanics 
as long as non-relativistic approximation is valid. 
Compared to the second quantization formalism, it is less convenient 
to calculate cross section, decay rate, branching ratio, and so on, 
due to the difficulty of formulating relativistic Brownian motion \cite{Dunkel}. 
However, when one considers a relationship between quantum 
and classical mechanics, above quantization scheme will be useful. 
As it is easy to contain classical equations, we can predict 
intermediate states between classical and quantum solutions 
to check the method.

Note, as we focus on low energy physics, 
the system with a single component function 
is sufficient for the description of dynamics. For the moment, we do not 
mention about a Hilbert space. 
In addition, instead of dealing with many particle system 
with the wave function, we can employ macroscopic physics 
by reinterpreting $\rho$ as the number density. 
In that case, Langevin equations should be modified 
but the same equations can be used with some simplification \cite{Nakamura}.

\section{Diffusion process}

It can be inferred from Eq. (\ref{MNE}) that the flip of the sign of $1-2s_1$ 
reproduces the diffusion equation. 
Only, it makes a wrong sign compared to the actual equation. 
To produce proper one, 
we set $f'=e^{\frac12\ln\rho-a'w}$ and 
$a'=D'/2D\sqrt{2s_1-1}$ with assuming $2s_1-1>0$, then
\begin{eqnarray}
2\sqrt{2s_1-1}mD\frac{\partial f'}{\partial t}
=2(2s_1-1)mD^2\nabla^2 f'
-2mD^2s_2(\nabla^2\ln\rho)f'+Vf',
\end{eqnarray}
When $s_2=0$ and $V=0$, it becomes
\begin{eqnarray}
\frac{\partial f'}{\partial t}
=\sqrt{2s_1-1}D\nabla^2 f'.
\end{eqnarray}
Thus $f'$ obeys the diffusion equation with the diffusion coefficient 
$D_0=\sqrt{2s_1-1} D$. This function is not directly related 
to the probability density so that we need an additional equation. 
To compute the dynamics, we consider the function 
$f=e^{\frac12\ln\rho+a'w}$ which satisfies
\begin{eqnarray}
\frac{\partial f}{\partial t}
=-\sqrt{2s_1-1}D\nabla^2 f.
\end{eqnarray}
Giving initial conditions of $f$ and $f'$, the probability density 
$\rho=ff'$ can be computed. With a famous solution in two dimensions, i.e. 
$f=f_0e^{r^2/4D_0(t+t_0)}/4\pi D_0(t+t_0)$ and 
$f'=f_0'e^{-r^2/4D_0(t+t_0')}/4\pi D_0(t+t_0')$, the probability density becomes 
\begin{eqnarray}
\rho
=\frac{t_0-t_0'}{4\pi D_0(t+t_0)(t+t_0')}
e^{-\frac{r^2(t_0-t_0')}{4D_0(t+t_0)(t+t_0')}}.
\end{eqnarray}
In this way, time evolution differs from the diffusion equation in general. 
For a special case, when $f$ is constant, the 
time-dependence becomes the same. 
The equations with the system of $f$ and $f'$ 
describe general dynamics of the diffusion equation. 

For the case $f=$ const. or $D'w=-\sqrt{2s_1-1}D\ln\rho$, 
the probability density obeys the diffusion equation. 
This case means $\bm v$ is proportional to 
$\bm u$, actually $\bm v=-\sqrt{2s_1-1}\bm u$. 
Suppose a solution $\rho=e^{-r^2/4D_0t}/4\pi D_0t$, 
we can calculate the friction from Eq. (\ref{vis}). 
Since $\bm u=-\bm x/2\sqrt{2s_1-1}t$, 
Eq. (\ref{vis}) reduces to $\beta=-(s_1-1)/\sqrt{2s_1-1}t$. 
This is one prediction for a particle obeying the diffusion equation in our system. 
It can be rewritten by observable parameters as 
\begin{eqnarray}
\beta=-2D_0W(-\pi r^2\rho)
\frac{\bm u^2-\bm v^2}{r^2|\bm u||\bm v|},
\end{eqnarray}
where $W(x)$ is the Lambert W-function. 
For large $t$ or small $\rho$, it can be approximated by 
$\beta\approx 2\pi D_0\rho
(\bm u^2-\bm v^2)/|\bm u||\bm v|$.
The sign of the friction force depends on magnitudes of two velocities; 
it is positive when $|\bm u|>|\bm v|$ and negative when $|\bm u|<|\bm v|$. 
For general $s_2$, the modified Nelson equations do not 
reduce to single equation in the above scheme. 
It becomes much harder to solve the equations.

\section{Soliton}
We have analyzed $s_1>1/2$ and $s_1<1/2$, and 
this section deals with $s_1=1/2$. 
The case $s_1=1/2$ and $s_2=0$ illustrate just classical physics. 
This is natural because this limit corresponds to 
$\hbar\rightarrow0$. Under the words of stochastic process, 
these assumptions lead 
$\frac12(d_++d_-)\bm b_\pm=\pm\beta \bm v+\bm K+d\bm B$. 
Then we get 
$(d_++d_-)\bm v=\bm K+d\bm B$
and $\frac12(d_++d_-)\bm u=\beta \bm v$. 
Since the average of 
$(d_++d_-)/dt$ corresponds to the Euler-Lagrange 
derivative, ordinary equations of motion realize. 
Precisely, they can be 
\begin{eqnarray}
(\frac{\partial}{\partial t}+\bm v\cdot\nabla)\bm v=\bm K,
\quad
(\frac{\partial}{\partial t}+\bm v\cdot\nabla)\bm u=\beta\bm v.
\end{eqnarray}
The first equation can lead Newton's equation of motion or  
it is related to the Navier-Stokes equations 
(to derive them we need to treat fluid density properly, see \cite{Cresson,Harvey}). 
The second equation is not 
trivial so it is a constrained system. Using the Fokker-Planck equations, 
it can be $(\bm u\cdot\nabla)\bm v+\beta \bm v=0$.

For the Hamiltonian with zero current-velocity, 
the condition $s_1=1/2$ makes
\begin{eqnarray}
H=-2s_2mD^2\nabla^2\ln\rho+V.
\end{eqnarray}
When the total energy and the potential are constant 
with written by $E-V=\Delta E$, it can be
\begin{eqnarray}
\Delta E=-2s_2mD^2\nabla^2\ln\rho.
\end{eqnarray}
In this section, we assume $s_2>0$ and $\Delta E>0$. 
For one spatial dimension, the probability density becomes Gaussian. 
For two dimensions with circular symmetry, 
the solution is $\rho=(c_1+c_2 r)e^{-\Delta E r^2/4s_2mD^2}$, 
where $c_1$ and $c_2$ are constant. 
For three dimensions with spherical symmetry, 
it is solved as
\begin{eqnarray}
\rho=
c_1 e^{-\frac{\Delta Er^2}{12s_2mD^2}-c_2/r}.
\end{eqnarray}
When $c_2>0$, it can be normalized by  
\begin{eqnarray}
c_1^{-1}=\frac14 \sqrt{\pi} c_2^3 G_{0,3}^{3,0}
 \left( \left. \frac{c_2^2 \Delta E}{48 s_2m D^2} \right|
-\frac32,-1,0 \right) ,
\end{eqnarray}
where 
$G$ is the Meijer G-function. 
When the current velocity is non-zero constant, we have
\begin{eqnarray}
E-\frac12m\bm v^2-V
=-\frac{2s_2mD^2}{4}\nabla^2\ln\rho,
\quad
\frac{\partial\ln\rho}{\partial t}
=-\bm v\cdot\nabla\ln\rho.
\end{eqnarray}
Writing $E-\frac12m\bm v^2-V=\Delta E$, 
the solution can be written in the same way 
with $\rho(t,\bm x+\bm v t)$: 
\begin{eqnarray}
\rho=
c_1 e^{-\frac{\Delta E(\bm x+\bm vt)^2}{12s_2mD^2}
-c_2/\sqrt{(\bm x+\bm vt)^2}}.
\end{eqnarray}
This solution contains both of Gaussian soliton and 
bubble soliton. When $c_2=0$ it is a Gaussian 
and otherwise it can be seen as a bubble. 
The size of a bubble is  
$r_0=(6s_2c_2mD^2/\Delta E)^{1/3}$ and 
the thickness of a film mainly depends on 
$\Delta r=\sqrt{12s_2mD^2/\Delta E}$. 
The soliton will be stable as it is a general solution in that environment. 
By changing $s_1$ and $s_2$, 
the soliton will collapse in several ways. 

More generally, when $\bm v=D'\nabla w$, 
$f=e^{\frac12\ln\rho+ibw}$, and 
$b=D'/\sqrt{8s_2D^2}$, 
we have
\begin{eqnarray}
\nabla^2 f
=(\frac14(\nabla\ln f^*f)^2
+\frac{D'}{4s_2D^2}\frac{\partial w}{\partial t}
+\frac{V}{4s_2mD^2}
-\frac{ib}{D'}\frac{\partial \ln\rho}{\partial t}
)f,
\end{eqnarray}
where $*$ denotes complex conjugate. 
Then we get
\begin{eqnarray}
i\sqrt{8s_2}mD\frac{\partial f}{\partial t}
=-4s_2mD^2\nabla^2 f
+(s_2mD^2(\nabla\ln f^*f)^2
+{V})f.
\end{eqnarray}
Thus the equations reduce to one differential equation 
for complex function $f$ as in the same way of 
the Schr\"odinger equation. 
This equation will generalize the above soliton solution.

\section{Periodic localization}

The Schr\"odinger equation appears by taking 
$s_2=0$ and $\rho=f^2$ in the Hamiltonian system. 
However, even when $s_2\not=0$, 
we can get the same equation. 
By reparametrizing the function 
$\rho=f^{2(1-2s_1+2s_2)/(1-2s_1)}$, 
the constant Hamiltonian without current 
velocity becomes 
\begin{eqnarray}
\label{mene}
E
=-\frac{2(1-2s_1+2s_2)^2mD^2}{1-2s_1}
\frac{\nabla^2 f}{f}+V.
\end{eqnarray}
If $1-2s_1>0$ and $E-V>0$, a typical solution is 
\begin{eqnarray}
f=c_0
\left|\cos[\frac{x_1}{\Delta x_1}+c_1]\right|
\left|\cos[\frac{x_2}{\Delta x_2}+c_2]\right|
\left|\cos[\frac{x_3}{\Delta x_3}+c_3]\right|.
\end{eqnarray}
For simplicity, we choose $c_1=c_2=c_3=0$ 
and $\Delta x_i=\Delta x$, then
\begin{eqnarray}
\Delta x=\sqrt{\frac{6(1-2s_1+2s_2)^2mD^2}{(1-2s_1)(E-V)}}.
\end{eqnarray}
In the view of probability density, it becomes
\begin{eqnarray}
\label{den}
\rho=\rho_0
\left|\cos[\frac{x_1}{\Delta x}]\right|^{\frac{1-2s_1}{2(1-2s_1+2s_2)}}
\left|\cos[\frac{x_2}{\Delta x}]\right|^{\frac{1-2s_1}{2(1-2s_1+2s_2)}}
\left|\cos[\frac{x_3}{\Delta x}]\right|^{\frac{1-2s_1}{2(1-2s_1+2s_2)}}.
\end{eqnarray}
If we take $\rho$ as the number density, 
$(1-2s_1)/2(1-2s_1+2s_2)\gg1$, and 
the Born-Oppenheimer  approximation, 
it can roughly describe nuclei of solid state. 
Suppose the diameter of a nucleus is 10fm 
and the size of atom is 100pm, 
a realistic model can be obtained in the vicinity of
$(1-2s_1)/2(1-2s_1+2s_2)\sim10^8$. 
If $s_1$ is exactly equal to $s_2$, 
strong localization is occurred by making 
the absolute values of $s_1$ and $s_2$ very large. 
In other cases, it needs fine-tuning to keep $1-2s_1+2s_2$ 
very small for the realization of such localization. 

Naively thinking, if it is regarded as a solid, 
phase transition can be performed in the framework of 
equations of motion. Varying the 
parameter $s_1$ to $s_1>1/2$, it satisfies the 
diffusion equation so that nuclei in the solid will diffuse. 
Considering the diffusion process describe liquid state, 
this procedure can be interpreted as solid to liquid phase transition. 
Depending on how $s_1$ changes, whole dynamics of 
this phase transition can be predicted. 

\begin{table}
\begin{center}
\begin{tabular}{|c|ccc|}
\hline
equation (solution) & $s_1$ & $s_2$ & $f$ \\
\hline
quantum & $s_1=1/2-\hbar^2/8m^2D^2$ & $s_2=0$ &complex\\
diffusion & $s_1>1/2$ & $s_2=0$ & real\\
classical  & $s_1=1/2$ & $s_2=0$ &real\\
soliton  & $s_1=1/2$ & $s_2>0$ &real \\
periodic localization  & $s_1<1/2$ & $s_2\not=0$ & real\\
NLS & $s_1=1/2-\hbar^2/8m^2D^2$ & $s_2\not=0$ & complex\\
\hline
\end{tabular}
\caption{Appearance of interesting equations and solutions 
are shown, depending on $s_1$, $s_2$ and $f$. 
If one believes that Planck constant is constant rigorously 
in quantum phenomena, 
$s_1$ should retain $1/2-\hbar^2/8m^2D^2$. 
However, if this is not true at some macroscopic scale, 
it can be any value within $s_1>1/2$.}
\end{center}
\end{table}

Next, let us consider generalization of above 
solution with non-zero current velocity. 
If we take $s_1=1/2-\hbar^2/8m^2D^2$, 
$s_2\not=0$, and $f=e^{\frac12 \ln\rho+iaw}$, 
a nonlinear term appears in the Schr\"odinger equation:
\begin{eqnarray}
\label{NSL}
i\hbar \frac{\partial f}{\partial t}
=-\frac{\hbar^2}{2m}\nabla^2 f
-\frac{s_2\hbar^2}{2(1-2s_1)m}(\nabla^2\ln f^*f)f
+Vf.
\end{eqnarray}
When the current velocity is zero and $\hbar$ is free, 
above periodic localized solution is appeared. 
The nonlinear term vanishes when the function is the plane wave, 
otherwise it affects to the energy at the same order of 
the kinetic term. By extending to a Hilbert space, 
the nonlinear term exists in each component, i.e. 
\begin{eqnarray}
i\hbar \frac{\partial f_k}{\partial t}
=-\frac{\hbar^2}{2m}\nabla^2 f_k
-\frac{s_2\hbar^2}{2(1-2s_1)m}(\nabla^2\ln f_k^*f_k)f_k
+{\hat V}f_k,
\end{eqnarray}
where $k$ is the index of the Hilbert space. 
Therefore, our nonlinear extension is quite different 
from ordinary interactions of quantum mechanics. 

Table 1 summarizes the equations and solutions 
that we have considered. 
In general, we need to solve nonlinear 
differential equations (\ref{MNS}), but when 
$1-2s_1\geq 0$ and $\bm v=D'\nabla w$, 
they reduce to one complex equation. 
For the case $1-2s_1<0$ and $s_2=0$, 
the equations become two independent 
diffusion equations about $f$ and $f'$ to 
give the density $\rho=ff'$. 
A remaining region ($1-2s_1<0$ and 
$s_2\not=0$) is much more difficult to be solved since the equations 
cannot reduce to single equation.

\section{Conclusion}

We have proposed a general framework 
that includes the Schr\"odinger equation, 
the diffusion equation, the soliton equation, and the Newton's equations of motion. 
Based on the modified Langevin equations 
for non-differentiable processes, the system can 
describe various motions depending on two parameters 
$s_1$ and $s_2$. Keeping $s_2=0$, the parameter 
$s_1$ can connect quantum, classical, and diffusion equations. 
On the other hand, taking non-zero value for $s_2$, 
there exist some interesting solutions like soliton and 
periodic localization. 

Since the framework can provide a lot of predictions, 
the confirmation of the theory seems possible. 
For instance, it includes generalized diffusion equations 
with the system of $\rho=ff'$ and the 
new type of soliton equation with the potential 
$\nabla^2\ln\rho$. Concretely, 
the potential terms $\bm u^2$ and $\nabla\cdot\bm u$ 
with arbitrary coefficients are the prediction of the theory. 
If one finds such potential in soliton, diffusion, 
and quantum states, these states can be related in our simple model.

\vspace{1cm}
\noindent
{\bf Acknowledgement}

The author is supported by Grand-in-Aid for Scientific Research,
No.23.696 from the Japan Society of Promotion of Science.

\end{document}